\documentstyle[12pt]{article}
\title{
\phantom. \hfill{$\scriptstyle {IFUM\ -\ 438/FT}$} \\
\phantom. \hfill{$\scriptstyle {March\ 1993}$}\\
\phantom. \hfill\\
Functional integration method\\
for 1D localization, multipoint correlators\\
and persistent current in mesoscopic ring\\
at arbitrary magnetic fields}
\author{{\sl I.V.Kolokolov}\\ Budker Institute of Nuclear
Physics,\\ Novosibirsk 630090, Russia} \date{}

\begin{document}
\maketitle

\begin{abstract}
Starting from the Abrikosov-Ryzhkin formulation of the
1D random potential problem I find closed functional representations
for various physical quantities. These functional integrals are
calculated exactly without the use of any perturbative expansions.
The expressions for the multipoint densities correlators are
obtained. Then I evaluate the mean square dispersion of the size
of localized wave functions. As a physical application of the
method, I find the expectation value of the persistent current
in mesoscopic ring with arbitrary magnetic flux $\Phi$. (For
small $\Phi$ this problem has been solved by O.Dorokhov). The
case when the random potential has finite correlation length
is considered too.
\end{abstract}

\newpage
\maketitle

\section{Introduction and definition of the model.}

   Anderson localization is acknowledged to be a fundamental
macroscopic quantum phenomena. The localization manifests itself most
evidently in one dimensional case. The essence of the effect consists
in all the eigenfunction of the Hamiltonian
\begin{equation}\label{Hamiltonian}
\hat{\cal H}=-\frac{d^2}{dx^2}+U(x),
\end{equation}
to be localized wave packets providing the potential  $U(x)$
is a random function of $x$.
(See the rigorous formulations, detailed discussion and bibliography
in the book \cite{Lifshits}.) This statement remains valid
in the high energy limit considered in the present paper.

The only quantities that can be
calculated directly are various averages over
an ensemble of potentials $U(x)$. The measure of this averaging is
reconstructed from the space correlation properties of a sample at our
disposal. In the simplest case of the white noise statistics it takes
the form:
\begin{equation}\label{Noise}
{\cal D}U\exp\left(-\frac{1}{2D}\int\limits_{-L}^{L}U^2(x)\,dx\right),
\end{equation}
\[
<U(x)U(x^\prime)>=D\delta (x-x^\prime).
\]

Here $(-L,L)$ is the interval, which our system takes up.
P.Anderson has shown \cite{Anderson} that the difference from zero
of the density-density correlator can be used
as a criterion of localization of the state with energy
$E$. The correlator can be expressed as
\begin{eqnarray}\label{Density}
p_E(x,x^\prime)&=\lim\limits_{L\to\infty}
<\sum\limits_n\delta(E-E_n)|\Psi_n(x)|^2
|\Psi_n(x^\prime)|^2>=\nonumber\\
&=\lim\limits_{L\to\infty}\lim\limits_{\epsilon\to
+0}\frac{\epsilon}{\pi} <|G(x,x^\prime|E+i\epsilon)|^2>,
\end{eqnarray}
where $\Psi_n(x)$ are the eigenfunctions of $\hat{\cal H}$:
\[
\hat{\cal H}\Psi_n(x)=E_n\Psi_n(x)
\]
and $G(x,x^\prime|E+i\epsilon)$ is the resolvent of $\hat{\cal H}$:
\begin{equation}\label{Resolvent}
\left(\hat{\cal H}-E\right)G(x,x^\prime|E+i\epsilon)=\delta (x-x^\prime),
\end{equation}

Indeed, the continuous spectrum wave functions are of the order of
$1/L^{1/2}$ in every space point,  and the sum over
$n$ gives effectively the factor $L$, so
$p_E(x,x^\prime)\sim 1/L\to 0$. For a homogeneous in the
average potential the probability to find the state
localized about
a given point is $\sim 1/L$, but the wave
function $\Psi_n(x)$ on its own does not depend asymptotically on
$L$ here. Hence only normalizable states contribute to
$p_E(x,x^\prime)$ in the thermodynamic limit.
(Assume boundary conditions in the endpoints of the interval $(-L,L)$
provide the hermiticity of $\hat{\cal H}$.)

To calculate quantities like (\ref{Density}) two approaches have been
developed. The first one is so called "phase formalism." It allows one, in
principle, to derive partial differential equations of the
Fokker---Plank kind for various averages over the ensemble (\ref{Noise})
\cite{Galperin,Frish}.
(See for the review of advances
\cite{Lifshits}.) However, such an approach gives explicit results
only in the regime:
\begin{equation}\label{Regime}
\frac{D}{E^{3/2}}\ll 1,
\end{equation}
corresponding to the quasiclassical kinetics.
Then another method can be used \cite{Ber}: extraction and summation
of the infrared-singular terms of perturbation theory series
(for review see  \cite{Gog}).
The direct performance of this program requires sophisticated
constructions and tedious computations.

It has been noted in \cite{Abr,Ryzh} that the sum of leading terms
in the above mentioned perturbation theory
corresponds to
some expectation values for spin $1/2$ placed in a random magnetic field
with Gaussian statistics.
\footnote {Conceptually the same method was used
in the paper \cite{Dor}.}

We present here  derivation of this spin model somewhat modifying
the  line of arguments.

For the wave function of the particle we assume the following
boundary conditions:
\begin{equation}\label{BCon}
\frac{d}{dx}\Psi_n(x=-L)=\Psi_n(x=L)=0.
\end{equation}

The Green function (\ref{Resolvent}) can be expressed in terms
of the solutions $u(x),\tilde u(x)$ of the initial-value problems:
\begin{equation}\label{Coshi}
\left(\hat{\cal H}-E\right)u=\left(\hat{\cal H}-E\right)\tilde
u=0,
\end{equation}
\[
u^\prime(-L)=0,\; u(-L)=1,\;
{\tilde u}^\prime (L)=1,\; \tilde u(L)=0;
\]
\[
G(x,x^\prime)=\frac{1}{W}
\left\{\matrix{
u(x)\tilde u(x^\prime), &x<x^\prime\cr
u(x^\prime)\tilde u(x), &x^\prime<x}\right.
\]
Here $W$ is the Wronskian of the functions $u$ and $\tilde u$:
\begin{equation}\label{Wronsky}
W=-u^\prime(x)\tilde u(x)+u(x)\tilde u^\prime(x).
\end{equation}
All the physical quantities of interest can be defined through one
solution only, say, $u(x)$ (see below). One can introduce for the
function $u(x)$ the "plane-wave components"
$v_1(x)$ and $v_2(x)$:
\[
v_1(x)=e^{-ikx}\left(u^\prime(x)+iku(x)\right)
\]
\begin{equation}\label{waves}
v_2(x)=-e^{ikx}\left(u^\prime(x)-iku(x)\right), E=k^2
\end{equation}
\[
u(x)=\frac{1}{2ik}\left(v_1(x)e^{ikx}+v_2(x)e^{-ikx}\right),
\]
so that $v_1 = 0$ ($v_2 = 0$) for the plane wave propagating from right
to left (from left to right). The equation (\ref{Coshi}) is equivalent
to the following first-order matrix equation:
\begin{equation}\label{matrequ}
\frac{d}{dx}\left(\begin{array}{c}v_1(x)\\v_2(x)\end{array}\right)
=\left(\begin{array}{cr}U(x)/2ik,&U(x)e^{-2ikx}/2ik\\
              -U(x)e^{2ikx}/2ik,&-U(x)/2ik\\
\end{array}\right)
\left(\begin{array}{c}v_1(x)\\v_2(x)\end{array}\right)
\end{equation}
and reduction
\begin{equation}\label{redu}
-v_1(x)=v_2^*(x).
\end{equation}
It is seen from (\ref{matrequ}) that the derivatives $v_1$ and $v_2$
with respect to $x$ are small along with the potential $U(x)$. That
is, $v_1(x)$  and $v_2(x)$ are changed slowly compared to $\exp(\pm ikx)$.
Let us rewrite
(\ref{matrequ}) in more compact notations:
\begin{equation}\label{dot}
\dot{\hat{v}}=\left(i\varphi(x)s^z + \zeta^+(x)s^- + \zeta^-(x)s^+
\right)\hat{v}.
\end{equation}
Here
\[
\hat{v}=\left(\begin{array}{c}v_1(x)\\v_2(x)\end{array}\right)
\]
\begin{equation}\label{denote}
\varphi(x)=-U(x)/k,\; \zeta^{\pm}(x)=\pm iU(x)\exp(\pm 2ikx)/2k,
\end{equation}
$s^z=\sigma^z/2, s^{\pm}=(\sigma^x \pm i\sigma^y)/2$ are
the usual spin operators and the dot denotes here and below the
$x$-derivative. The formal solution of
(\ref{dot}) can be written in the form:
\[
\hat{v}(x)={\cal T}(x,-L)\hat{v}(-L)
\]
\begin{equation}\label{Texp}
{\cal T}(x,-L) = T\exp \left(\int\limits_{-L}^{x}(i\varphi(t)s^z+
\zeta^+(t)s^-+\zeta^-(t)s^+)\,dt\right),
\end{equation}
where the sign $
$ implies the product is ordered along the interval
$(-L,L)$.

Let us consider the expectation value of some functional of
$v_1(x)$,$v_2(x)$. Expanding the $
$-exponential (\ref{Texp}) and this
functional in a series in the fields $\varphi(t),\zeta^{\pm}(t)$
and performing the averaging over ${\cal D}U(x)$  we obtain the result
as a series in integrals:
\[
\int dt\,dt^\prime<\varphi(t)\varphi(t^\prime)>, \;
\int dt\,dt^\prime<\zeta^+(t)\zeta^-(t^\prime)>,
\]
\begin{equation}\label{osc}
\int dt\,dt^\prime<\varphi(t)\zeta^{\pm}(t^\prime)>, \;
\int dt\,dt^\prime<\zeta^+(t)\zeta^+(t^\prime)>, \;
\int dt\,dt^\prime<\zeta^-(t)\zeta^-(t^\prime)>, \;
\end{equation}
over some domains of the order of $L$.

In the last three expressions we integrate fast-oscillating functions.
Therefore these integrals remain restricted in their values
with increasing integration intervals and fall with an increase of
energy. On their turn, integrals of the first two kinds correspond to the
infrared-singular contributions and grow linearly with $L$.
Thus to leave in the perturbation theory series the terms
dominating in the large $L$ limit the correlators:
$<\zeta^+\zeta^+>,
<\zeta^-\zeta^->$, and $<\varphi,\zeta^{\pm}>$ should be
neglected.
It is equivalent to the assumption that the fields
$\varphi$ and $\zeta^{\pm}$  are statistically independent and the
weight of ${\cal D}\zeta^{\pm}$ - averaging is phase invariant.
For the white noise statistics the corresponding integration measure
has the form:
\begin{equation}\label{measure}
{\cal D}\varphi(x){\cal D}\zeta^{\pm}(x)
\exp\left\{-\frac{2}{\alpha}\int\limits_{-L}^{L}\left(a\varphi^2(x)+
\zeta^+(x)\zeta^-(x)\right)\,dx\right\},
\end{equation}
where
\begin{equation}\label{alpha}
\alpha=\frac{D}{2k^2}, \; a=\frac{1}{8}.
\end{equation}
We shall consider below the parameter $a$ as an arbitrary ones. (It
does not enter the final results.)

The formulae (\ref{Texp})
and (\ref{measure}) were first proposed for the one dimensional
random potential problem in the work \cite{Abr}. Our
presentation of it does not refer to the existence of the Fermi
level. It allows one to suppose that Abrikosov-Ryzhkin model
has some universal features relevant to the infrared behaviour.
The model can be easily generalized to random potentials with finite
correlation length. It may be usable to study spectral properties of
operators which are not random in the strict sense (see Conclusion).

The terms neglected in deriving of (\ref{Texp}), (\ref{measure})
are smaller by a factor
$\sim 1/(kL)$ compared to the ones kept. Thus this model can be
applied to the study of mesoscopic systems (see section 5), since
the inequality $1/(kL)\ll 1$ for sufficiently large $k$
is compatible with $l\geq L$,
where $l$ is the mean free path.

The authors of the paper \cite{Abr} have used
the formulae (\ref{Texp}), (\ref{measure})
to obtain the conductivity of a one dimensional metal.
Unfortunately, the calculations have being carried out there by
perturbation theory method lead to cumbersome
constructions, which are inadequate to simple model.
In the present paper I solve this Abrikosov-Ryzhkin model
exactly with the help of functional integration method.
On deriving the path integral representation I find the
multipoint correlators of arbitrary powers of the density. With the
use of these expressions I evaluate the mean-square dispersion
of the size of localized wave function. As a physical application
of the method I calculate the mean absolute value of the persistent current
in a mesoscopic ring with an arbitrary magnetic flux
$\Phi$. (For small $\Phi$ it has been
found recently in \cite{Dor}.)
In Conclusion I analyze the localization length dependence on the
correlation length of the random potential. I discuss also
a quantity that
could play the role of the order parameter describing
localization.

\section{Functional representation for averaged
functionals of $\hat v(x)$}

It is impossible to express $\hat v(x)$ as a functional of the
fields $\varphi(x)$, $\zeta^{\pm}(x)$ explicitly. The same
problem arises when one undertakes an attempt to write out a closed functional
representation for the partition function of quantum Heisenberg
ferromagnet. It has been solved in the works
\cite{Kol1}-\cite{Kol3} and here we take advantage of the method
proposed there.

The ordered exponential ${\cal T}(x,-L)$  is defined by the equation
\begin{equation}\label{Teq}
\dot{\cal T} =  (i\varphi(t)s^z+
\zeta^+(t)s^-+\zeta^-(t)s^+){\cal T}
\end{equation}
and the initial condition:
\begin{equation}\label{Tinit}
{\cal T}(x=-L,-L)=1.
\end{equation}
Let us consider the operator given as a product of usual matrix exponential:
\begin{equation}\label{Anzatz}
\tilde {\cal T}(x,-L)=\exp \left(s^+\psi^-(x)\right)
\exp \left(is^z\int\limits_{-L}^{x} \rho\,dt\right)
\exp \left(s^-\int\limits_{-L}^{x}\,dt\psi^+(t)\exp\left(i
\int \limits_{-L}^{t} \rho\,d\tau\right)\right)\times
\end{equation}
\[
\times \exp \left(-s^+\psi^-(-L)\right).
\]
Here $\psi^{\pm}(x), \rho(x)$ are some new fields.
It obeys the equation:
\begin{equation}\label{Anzeq}
\dot{\tilde {\cal T}}=\left\{(i\rho + 2\psi^+\psi^-)s^z+
\psi^+s^- + (\dot{\psi^-} - i\rho\psi^- - \psi^+(\psi^-)^2)s^+\right\}
\tilde {\cal T},
\end{equation}
and the last factor in (\ref{Anzatz}) gives equality:
\begin{equation}\label{Anzini}
\tilde {\cal T}(-L,-L)=1.
\end{equation}
Thus, the change of variables in the functional integral over the
measure
(\ref{measure}):
\[
i\varphi=i\rho+2\psi^+\psi^-,
\]
\begin{equation}\label{Change}
\zeta^- = \dot{\psi^-} - i\rho\psi^- - \psi^+(\psi^-)^2
\end{equation}
\[
\zeta^+=\psi^+
\]
brings the ordered exponential ${\cal T}(x,-L)$ to the form (\ref{Anzatz}):
\begin{equation}\label{Anzz}
{\cal T}(x,-L)=\tilde {\cal T}(x,-L),
\end{equation}
and allows us to obtain an explicit  functional integral
representation for any physical quantity to be averaged.
(Parametrization of $SL(2,C)$-valued functions
on two-dimensional space analogous to the (\ref{Change})
has been used also in the paper \cite{Ger}).
To accomplish the change of variables in functional
integral, much like the usual integrals,
we need know the map (\ref{Change}) in one direction only:
from $(\phi,\zeta^{\pm})$ to $(\rho,\psi^{\pm})$.
The Jacobian ${\cal J}[\rho,\psi^{\pm}]$:
\begin{equation}\label{Jacdef}
{\cal D}\varphi{\cal D}\zeta^+ {\cal D}\zeta^- =
{\cal J}[\rho,\psi^{\pm}]{\cal D}\rho{\cal D}\psi^+ {\cal D}\psi^-
\end{equation}
depends on the regularization of the map (\ref{Change})
and on the kind of condition imposed on the field $\psi^-$.
The latter is necessary since there is first-order derivative
of $\psi^-$ on the right-hand side of (\ref{Change}).
The periodic boundary condition
renders the map (\ref{Change}) irreversible. Following
papers \cite{Kol2}, \cite{Kol3} we consider the field $\psi^-(x)$
as obeying an initial condition:
\begin{equation}\label{Inc}
\psi^-(-L)= \psi_0,
\end{equation}
but, unlike \cite{Kol2}, \cite{Kol3}, the concrete value of
$\psi_0$ will be picked as the situation requires.

The regularization of the map  (\ref{Change}) is determined by the
physical meaning of the model:
the white-noise correlator (\ref{Noise}) is to be considered
as the limit of a smooth symmetrical correlation function. Any such
a regularization of the
$\delta$-function gives for the correlators:
\begin{equation}\label{Cc}
\langle\zeta^+(t)\int\limits_{0}^{t}\zeta^-(t^\prime)\,dt^\prime
\rangle=\langle\zeta^-(t)\int\limits_{0}^{t}\zeta^+(t^\prime)\,dt^\prime
\rangle
\end{equation}
the limiting value equal to $\frac{1}{2}\frac{\alpha}{2}$ what
corresponds to the extension of a definition of the step function
$\theta(x)$:
\begin{equation}\label{Theta}
\theta(0)= 1/2.
\end{equation}
The discrete version of the
change of variables (\ref{Change}) providing the equalities
(\ref{Cc}) has the form:
$\left(\zeta_{n}^{\pm}=\zeta^{\pm}(t_{n}), \rho_{n}=\rho(t_{n}),\dots,
n=1,\dots,M, t_{n}=-L+\frac{2Ln}{M}, h=\frac{2L}{M}\rightarrow 0,
M\rightarrow\infty\right),$
\[
i\phi_{n}=i\rho_{n}+\psi_{n}^{+}(\psi_{n}^{-}+\psi_{n-1}^{-}),
\]
\begin{equation}\label{Regul}
\zeta_{n}^{-} = \frac{1}{h}(\psi_{n}^{-} - \psi_{n-1}^{-}) -
\frac{1}{2}i\rho_{n}(\psi_{n}^{-}+ \psi_{n-1}^{-}) -
\frac{1}{4}\psi_{n}^{+}(\psi_{n}^{-}+\psi_{n-1}^{-})^2,
\end{equation}
\[
\zeta_{n}^{+}=\psi_{n}^{+}.
\]
All the over-diagonal elements of the differential matrix
of the map (\ref{Regul}) equal  zero. Then the Jacobian
${\cal J}$, being the determinant of this matrix, is equal to the
product of the diagonal elements only:
\begin{equation}\label{Jacrez}
{\cal J}=
\hbox{const}\,\exp\left( -\frac{i}{2} \int\limits_{-L}^{L}\rho\,dt \right).
\end{equation}
Making the substitution (\ref{Change}) into the measure (\ref{measure})
with the use of the expressions (\ref{Jacdef}), (\ref{Jacrez})
we obtain the weight of averaging over the fields
$(\rho,\psi^{\pm})$:
\[
N{\cal D}\rho{\cal D}\psi^+ {\cal D}\psi^- \exp\left(-S(\rho,\psi^{\pm})
\right),
\]
\begin{equation}\label{Action}
S(\rho,\psi^{\pm})=\frac{2}{\alpha}\int\limits_{-L}^{L}\,dx
\left(a\rho^2+\psi^+\dot{\psi^-} - (1+4a)i\rho\psi^+\psi^-
-(1+4a)(\psi^+\psi^-)^2\right) +
\frac{i}{2}\int\limits_{-L}^{L}\,dx\rho.
\end{equation}
Here $N$ is a normalization constant depending on $\alpha L$.

In calculating of the Jacobian (\ref{Jacdef}) we were considering
$(\rho,\psi^{\pm})$ and  $(\phi,\zeta^{\pm})$ as sets of independent
complex variables or, in other words, as distinct coordinate systems in
the whole space ${\cal C}^{3M}$ of the fields' configurations.
The conditions
\begin{equation}\label{surface}
Im \varphi=0, \zeta^+=(\zeta^-)^*,
\end{equation}
being from the outset embedded in the model specify
the surface $\Sigma$ in ${\cal C}^{3M}$ along which the
differential form
${\cal D}\varphi\bigwedge {\cal D}\zeta^+ \bigwedge{\cal D}\zeta^-$
or
${\cal D}\rho\bigwedge{\cal D}\psi^+ \bigwedge{\cal D}\psi^-$
is integrated. From the point of view of the coordinates set
$(\rho,\psi^{\pm})$ the equation
(\ref{surface}) for $\Sigma$ is implicit. According to
the Cauchu-Poincare theorem the integration surface can be deformed
in an arbitrary way in the convergence domain while an analytical
function is integrated. There exists a continuous family of surfaces
(homotopy) situated as a unit in "perturbative" convergence domain,
which includes both the surfaces
$\Sigma$ and the "standard" one ${\Sigma}^\prime$:
\begin{equation}\label{ssurface}
{\Sigma}^\prime=\left\{Im \rho=0, \psi^+=(\psi^-)^*\right\}.
\end{equation}
The word "perturbative" means here that we check the convergence
in every order of the perturbation theory expansion. (This homotopy
is presented explicitly in the paper \cite{Kol1}.)
Thus, treating functional integral as the sum of
perturbation theory series \cite{Po} we can replace the surface
of integration $\Sigma$
by the standard one ${\Sigma}^\prime$.

However, to pass from $\Sigma$ to ${\Sigma}^\prime$
the expressions being averaged (and not just the action) should be
written in the form allowing the direct analytical continuation
from the surface $\Sigma$.
It means constructively that the definition of any physical quantity
in terms of the matrix elements of ${\cal T}(x,-L)$
must contain no complex conjugations.

\section{The density-density correlator expression in terms of
the functions $\hat v(x)$}

The formula (\ref{Density}) defines the correlator $p_E(x,x^\prime)$
in terms of the singular at $\epsilon\to +0$
part of the Green function $G(x,x^\prime|E+i\epsilon)$.
When we use the representation (\ref{Wronsky})
the singularity appears owing to zeros of the Wronskian $W(E)$
on the real axis. Neglecting
$\epsilon$ in the numerator (\ref{Wronsky}) and substituting
\begin{equation}\label{Wronskexp}
W(E \pm i\epsilon)=W(E) \pm i\epsilon W^\prime(E)
\end{equation}
into the denominator we obtain:
\begin{equation}\label{Ageden}
p_E(x,x^\prime)=\left\langle\frac{u^2(x)\tilde
u^2(x^\prime)}{|W^\prime(E)|} \delta(W)\right\rangle, x^\prime >x.
\end{equation}
$W(E)$ does not depend on $x$ and, thus, we may put in
(\ref{Wronsky}) $x=L$:
\begin{equation}\label{WronskL}
W= u(L)
\end{equation}
Being in the product with $\delta (W) = \delta \left(u(L)\right)$
the solution $\tilde u(x)$ is proportional to $u(x)$. The
proportionality coefficient is determined by the
conditions (\ref{Coshi}).
So, for an arbitrary functional
${\cal F}\left[\tilde u(x)\right]$ the following equality
takes place:
\begin{equation}\label{Funk}
{\cal F}\left[\tilde u(x)\right]\delta \left(u(L)\right)=
{\cal F}\left[\frac{u(x)}{u^\prime(L)}\right]\delta \left(u(L)\right).
\end{equation}
According to (\ref{WronskL}) we can express $W^\prime(E)$ in terms
of the derivative of $u(x)$ with respect to the energy $E$:
\begin{equation}\label{Wropro}
W^\prime(E)=\frac{\partial u(L)}{\partial E}.
\end{equation}
The function $g(x)=\partial u(x)/\partial E $ obeys the equation
\begin{equation}\label{g-eq}
\left(-\frac{d^2}{dx^2}+U(x)-E\right)g(x)=u(x),
\end{equation}
and the initial condition:
\begin{equation}\label{g-eqin}
g(x=-L)=g^\prime (x=-L)=0.
\end{equation}
The substitution $g(x)=q(x)u(x)$ leads to the first-order
equation for $q^\prime(x)$; its solution gives us:
\begin{equation}\label{g-sol}
g(x)=u(x)\int\limits_{-L}^{x}\,\frac{dy}{u^2(y)}
\int\limits_{-L}^{y}\,dy_{1} u^2(y_{1}),
\end{equation}
and:
\begin{equation}\label{InWro}
\frac{1}{|W^\prime(E)|}\delta\left(u(L)\right)=
\frac{1}{|g(L)|}\delta\left(u(L)\right)=
\frac{|u^\prime(L)|}{\int\limits_{-L}^{L}u^2(y)\,dy}\delta\left(u(L)\right).
\end{equation}
Thus the correlator $p_E(x,x^\prime)$ can be written via
$u(x)$ as follows:
\begin{equation}\label{den-cosh}
p_E(x,x^\prime)=\left\langle\frac{u^2(x)
u^2(x^\prime)}{|u^\prime(L)|\int\limits_{-L}^{L}u^2(y)\,dy}
\delta\left(u(L)\right)\right\rangle, x^\prime >x.
\end{equation}
In the high-energy limit (\ref{Regime}) we can get rid of the
$\delta$-function and obtain a simple formula for $p_E(x,x^\prime)$
in terms of slowly varying amplitudes $v_{1,2}(x)$. Indeed, in
neighbourhood of any given point
$x_0$ the function $u(x)$ can be written in the form:
\begin{equation}\label{u-sinus}
u(x)=u_{sl}(x) \sin (kx+\delta)
\end{equation},
where the envelope $u_{sl}(x)$ and the phase $\delta$ vary only
slightly over distances of the order $\sim 1/k$.
Let us average the expression (\ref{den-cosh}) over the interval
$\Delta L$
of the right endpoints' positions of our "space" $(-L,L)$:
\begin{equation}\label{Averden}
\tilde p_E(x,x^\prime)=\frac{1}{\Delta L}
\int\limits_{L}^{L+\Delta L}p_E(x,x^\prime)\,dL
\end{equation}
\begin{equation}\label{Delta L}
\frac{1}{k} \ll \Delta L \ll \frac{2}{\alpha}\equiv l.
\end{equation}
(Here we introduce the standard notation $l$ for the localization length.) In
the thermodynamic limit the functions
$\tilde p_E(x,x^\prime)$ and $p_E(x,x^\prime)$ coincide.
On the other hand, the value of
$u(x)$ in a given point, by the construction, does not depend on
the right endpoints position. The integral in the denominator of
(\ref{den-cosh}) is determined by the envelope
$u_{sl}(x)$ only. The variation of $L$ from $L$ to $L+\Delta L$
does affect it asymptotically. The averaging
(\ref{Averden}) is sufficient only for the factor
$\delta\left(u(L)\right)/|u^\prime(L)|$. Since the conditions
(\ref{Delta L}) mean that $u_{sl}(x)$
can be considered as a constant in the averaging interval, we obtain:
\begin{equation}\label{Averdelt}
\frac{1}{\Delta L}
\int\limits_{L}^{L+\Delta L}dL\,
\frac{1}{|u^\prime(L)|}\delta\left(u(L)\right)=
\frac{1}{\pi k\,u_{sl}^2(L)},
\end{equation}
We can derive  similarly the relationship between
$u_{sl}^2(x)$ and $u^2(x)$, in particular:
\begin{equation}\label{Aversqu}
u_{sl}^2(L) \approx
\frac{2}{\Delta L}
\int\limits_{L}^{L+\Delta L}dL\,u^2(L).
\end{equation}
Substituting into (\ref{den-cosh})-(\ref{Aversqu})
the expression of $u(x)$ via $\hat v(x)$, neglecting the contributions
vanishing in the
$k\to \infty$ limit and keeping in the numerator of
(\ref{den-cosh}) the "resonance" terms only, we obtain:
\begin{equation}\label{den-v-expr}
p_E(x,x^\prime) \approx \tilde p_E(x,x^\prime) \approx \frac{1}{2\pi k}
\left\langle\frac{v_{1}(x)v_{2}(x)v_{1}(x^\prime)
v_{2}(x^\prime)}{v_{1}(L)v_{2}(L)\int\limits_{-L}^{L}v_{1}(y)v_{2}(y)\,dy}
\right\rangle, x^\prime >x.
\end{equation}
The "non-resonance" terms containing the oscillating factors
$\exp \pm 2ik(x-x^\prime)$ will result in exponentially small in
$\alpha L$ contributions and, thus, can be neglected.

\section{Functional integration for correlators of the
density -density type}

The form of the expression (\ref{den-v-expr}) allows direct analytical
continuation from the surface $v_{1}=(v_{2})^*$ over the functions
$v_{1,2}(x)$. It is not restrictive to put
$\exp (ikL)=1$ in the thermodynamic limit. (Only if the formula
(\ref{den-v-expr}) has been obtained already!) Initial condition
for $\hat v(x)$ takes the form:
\begin{equation}\label{Inc-v}
\hat{v}(-L)=ik\left(\begin{array}{c}1\\1\end{array}\right).
\end{equation}
To find $\hat v(x)$ we substitute into (\ref{Texp}) the expression
(\ref{Anzatz}) for the evolution operator
${\cal T}(x,-L)$ picking the quantity
$\psi_0$ to be equal to 1:
\begin{equation}\label{Inc-psi}
\psi_0=\psi^-(-L)=1.
\end{equation}
It yields the equality:
\begin{equation}\label{v-psi}
\hat{v}(x)=\exp \left(-\frac{i}{2}\int\limits_{-L}^{x} \rho\,dt\right)
\left(\begin{array}{c}\psi^-(x)\\1\end{array}\right),
\end{equation}
and the expression for $p_E(x,x^\prime)$:
\begin{equation}\label{den-psi-expr}
p_E(x,x^\prime) \approx \frac{1}{2\pi k}
\left\langle\frac{\psi^-(x)\psi^-(x^\prime)
\exp\left(-i\int\limits_{-L}^{x} \rho\,dt
-i\int\limits_{-L}^{x^\prime} \rho\,dt
+i\int\limits_{-L}^{L} \rho\,dt\right)}
{\psi^-(L)\int\limits_{-L}^{L}\psi^-(y)
\exp\left(-i\int\limits_{-L}^{y} \rho\,dt\right)\,dy}
\right\rangle, x^\prime >x.
\end{equation}
Here the averaging over ${\cal D}\rho{\cal D}\psi^+ {\cal D}\psi^- $
is carried out with the weight (\ref{Action}).
To calculate this functional integral we employ a trick similar to
the so-called "bosonization" in the field theory models
\cite{Pol}. Using the identity:
\[
\exp \left(-S(\rho,\psi^{\pm})\right)=
\int{\cal D}\eta \exp\left(-\tilde S(\eta,\rho,\psi^{\pm})\right),
\]
\begin{equation}\label{Bos-Action}
\tilde S(\eta,\rho,\psi^{\pm})=\frac{2}{\alpha}\int\limits_{-L}^{L}\,dx
\left((1+4a)\eta^2+a\rho^2+
\psi^+\dot{\psi^-} + (1+4a)(2\eta-i\rho)\psi^+\psi^-\right) +
\frac{i}{2}\int\limits_{-L}^{L}\,dx\rho,
\end{equation}
and the gauge transformation:
\begin{equation}\label{Gauge}
\psi^{\pm}(x)=\chi^{\pm}\exp\left(
\pm (1+4a)\int\limits_{-L}^{x}dt\,(2\eta-i\rho)
\right),
\end{equation}
we get rid of the non-linear terms in the action. The Jacobian
of the rotation (\ref{Gauge}) is equal to
\begin{equation}\label{Jacrot}
{\cal J}_{R}=const\,\exp\left( -\frac{1+4a}{2}
\int\limits_{-L}^{L}(2\eta-i\rho)\,dt \right),
\end{equation}
where the regularization (\ref{Regul}) is taken into account.
The fields $\eta$ and $\rho$ enter the
equation (\ref{den-psi-expr}) via the combination
\[
\int\limits_{-L}^{x}\left(2(1+4a)\eta-4ia\rho \right)\,dt
\]
only.
It is natural to consider it as a new integration variable:
\[
\dot{\xi}=2(1+4a)\eta-4ia\rho ,
\]
\begin{equation}\label{xi}
\xi(-L)=0
\end{equation}
\[
{\cal D}\rho{\cal D}\eta=const\,{\cal D}\rho{\cal D}\xi
\]
Then the Gaussian ${\cal D}\rho$-integration can be done easily and
we obtain the expressions for the measure:
\begin{equation}\label{xi-mes}
const\,{\cal D}\xi{\cal D}\chi^+ {\cal D}\chi^-
\exp \left(-\frac{1}{2\alpha}\int\limits_{-L}^{L}\,dx\dot{\xi}^2\;
-\frac{2}{\alpha}\int\limits_{-L}^{L}\,dx
\chi^+\dot{\chi^-}\; -\frac{\xi(L)}{2}\right)
\end{equation}
and for the quantity to be averaged:
\begin{equation}\label{den-xi-expr}
p_E(x,x^\prime) = \frac{1}{2\pi k}
\left\langle\frac{\chi^-(x)\chi^-(x^\prime)
\exp\left(-\xi(x)-\xi(x^\prime)+\xi(L)
\right)}
{\chi^-(L)\int\limits_{-L}^{L}\chi^-(y)
e^{-\xi(y)}\,dy}
\right\rangle, x^\prime >x,
\end{equation}
(The asymptotic equality in the limit (\ref{Regime}) is assumed.)
The initial condition for the field $\chi^-(x)$
follows from (\ref{Inc-psi}):
\begin{equation}\label{Inc-chi}
\chi^-(-L)=1.
\end{equation}
It means that $\chi^-(x)$ contains both the fluctuating part
$\chi_{f}^-(x)$ and the regular one:
\begin{equation}\label{chi-fluc}
\chi^-(x)=1+\chi_{f}^-(x),\; \chi_{f}^-(-L)=0.
\end{equation}
The component $\chi_{f}^-(x)$
does not contribute to $p_E(x,x^\prime)$ because the conjugated field
does not appear in the broken brackets in (\ref{den-xi-expr}).
Thus, the only averaging over the field $\xi(x)$ remains. Its
weight has the form
\begin{equation}\label{xi-weight}
\exp\left(-\frac{\alpha L}{4}\right)N^\prime\,
{\cal D}\xi
\exp \left(-\frac{1}{2\alpha}\int\limits_{-L}^{L}\,dx\dot{\xi}^2\;
-\frac{\xi(L)}{2}\right)
\end{equation}

Here the normalization constant $N^\prime$ is determined by the
quadratic in $\dot{\xi}$ term of the action:
\begin{equation}\label{norma}
N^\prime\,
{\cal D}\xi
\exp \left(-\frac{1}{2\alpha}\int\limits_{-L}^{L}\,dx\dot{\xi}^2\right)
=1.
\end{equation}
The factor $\exp\left(-\alpha L/4\right)$
provides the equality $<1>=1$ for the averaging over the entire
measure (\ref{xi-weight}). Thus we arrive at the following path
integral for the correlator
$p_E(x,x^\prime)$:
\begin{eqnarray}\label{Density-Integral}
&p_E(x,x^\prime)={\displaystyle \frac{1}{2\pi k}}
N^\prime
\exp\left(-\frac{\alpha L}{4}\right)
\int\limits_{\xi(-L)=0}{\cal D}\xi
\exp \left(-\frac{1}{2\alpha}\int\limits_{-L}^{L}\,dt\dot{\xi}^2
\,
-\frac{\xi(L)}{2}\right)\times
\nonumber\\
\!\!&\times
\exp\left(-\xi(x)-\xi(x^\prime)+\xi(L)\right)
\left\{\int\limits_{-L}^{L}
\exp\left(-\xi(t)\right)\,dt\right\}^{-1}
=\nonumber\\
&={\displaystyle \frac{N^\prime}{4\pi k\alpha}}
\exp\left(-{\displaystyle \frac{\alpha L}{4}}\right)
\int\limits_{0}^{\infty}d\lambda
\int\limits_{\xi(-L)=0}{\cal D}\xi
\exp \left(-\frac{1}{2\alpha}\int\limits_{-L}^{L}\,dt
\left(\dot{\xi}^2+\lambda e^{-\xi}\right)\;
+\frac{\xi(L)}{2}\right)
e^{-\xi(x)-\xi(x^\prime)}
=\nonumber\\
&=N^\prime\,(4\pi k\alpha)^{-1}\times\\
&\times\int\!\!\!
\int\limits_{-\infty}^{+\infty}d\sigma d\sigma^\prime
\exp\left(\frac{\sigma+\sigma^\prime}{2}\right)\!\!
\int\limits_{\xi(-L)=\sigma^\prime,\xi(L)=\sigma}\!\!{\cal D}\xi
\exp \left(-\frac{1}{2\alpha}\int\limits_{-L}^{L}dt
\left(\dot{\xi}^2+e^{-\xi}\right)-
{\displaystyle \frac{\alpha L}{4}}\right)
e^{-\xi(x)-\xi(x^\prime)}.
\nonumber
\end{eqnarray}
The last equality has been attained by changing of variables:
\begin{equation}\label{Shift}
\lambda=e^{-\sigma^\prime}, \xi\rightarrow\xi-\sigma^\prime
\end{equation}
and by separating the integrals over the values of $\xi(t)$
in the endpoints
$t=L$ and $t=-L$. The final path integral in (\ref{Density-Integral})
is of the Feynmann-Kac type
\cite{Feynmann} and it is equal to the following matrix
element:
\begin{eqnarray}\label{mx-el}
&p_E(x,x^\prime)=
\exp\left(-{\displaystyle \frac{\alpha L}{4}}\right)\,(4\pi k\alpha)^{-1}
\times\\
&\times
\langle e^{\xi/2}|\exp\left(-(L-x^\prime)\hat{H}\right)
e^{-\xi}\exp\left(-(x^\prime-x)\hat{H}\right)e^{-\xi}
\exp\left(-(x+L)\hat{H}\right)|e^{\xi/2}\rangle,
\nonumber
\end{eqnarray}
with the Hamiltonian:
\begin{equation}\label{Ham-eff}
\hat{H}=-\frac{\alpha}{2}\partial_{\xi}^{2}+\frac{1}{2\alpha}e^{-\xi}.
\end{equation}
The function $e^{\xi/2}$ increase when $\xi\rightarrow\infty$
and, consequently, it cannot be represented as a linear
combination of the eigenfunctions of $\hat{H}$:
\begin{equation}\label{Eigenfunctions}
\hat{H}\,f_{\nu}(\xi)=\frac{\alpha}{2}\nu^2 f_{\nu}(\xi),\;\;
f_{\nu}(\xi)=\frac{2}{\pi}\sqrt{\nu\, \sinh 2\pi\nu}\,
K_{2i\nu}\left(\frac{2}{\alpha}e^{-\xi/2}\right),
\end{equation}
\[
\langle f_{\nu}|f_{\nu^\prime}\rangle=\delta(\nu - \nu^\prime).
\]
Still, explicit solution of the corresponding evolution
equation leads to the following asymptotic relation:
\begin{equation}\label{Asymps}
\exp(-T \hat{H})\,
e^{\xi/2}\longrightarrow \exp\left(\frac{\alpha T}{8}\right)
\Upsilon_{0}(\xi)=\frac{2}{\alpha}\exp\left(\frac{\alpha T}{8}\right)
K_{1}\left(\frac{2}{\alpha}e^{-\xi/2}\right).
\end{equation}
Here $K_{\mu}(z)$ is the standard notation for the modified Bessel
function. The function
\[
\Upsilon_{0}(\xi)e^{-\xi}
=\frac{2}{\alpha}
K_{1}\left(\frac{2}{\alpha}e^{-\xi/2}\right)e^{-\xi}
\]
in its turn can be expanded in terms of
the complete set (\ref{Eigenfunctions}).
Thus, the correlator
$p_E(x,x^\prime)$ is equal to:
\begin{eqnarray}\label{p}
&p_E(x,x^\prime)={\displaystyle \frac{1}{4\pi k\alpha}}
\exp\left(-{\displaystyle \frac{\alpha |x-x^\prime|}{8}}\right)
\langle \Upsilon_{0}(\xi)e^{-\xi}|\exp\left(-|x-x^\prime|\hat{H}\right)
|\Upsilon_{0}(\xi)e^{-\xi}\rangle=\nonumber\\
&={\displaystyle \frac{\alpha}{\pi^3 k}}
\exp\left(-{\displaystyle \frac{\alpha |x-x^\prime|}{8}}\right)
\int\limits_{0}^{\infty}d\nu\,\nu\,\sinh 2\pi\nu
\exp\left(-\frac{\alpha \nu^2}{2} |x-x^\prime|\right)
\left\{ \int\limits_{0}^{\infty}dy\,y\,K_{1}(y)K_{2i\nu}(y)\right\}^2=
\nonumber\\
&={\displaystyle \frac{\alpha \pi}{2 k}}
\exp\left(-{\displaystyle \frac{\alpha |x-x^\prime|}{8}}\right)
\int\limits_{0}^{\infty}{\displaystyle
\frac{d\nu\,\nu\,\sinh\pi\nu}{\cosh^3\pi\nu}}
\left(\nu^2+\frac{1}{4}\right)^2
\exp\left(-\frac{\alpha \nu^2}{2} |x-x^\prime|\right)
\end{eqnarray}
The formula (\ref{p}) up to the redefinition
$\alpha/2=l^{-1}$ coincides with the well known
result \cite{Lifshits},\cite{Gog}.
This method allows us to compute the high-order correlators
as well. For example:
$(x_1<x_2<\dots<x_{2m}, m>1)$

\begin{eqnarray}\label{p-qm}
 & 2\pi k\, p^{(q,m)}_{E}(x_{1}, x_{2},\dots x_{2m-1}, x_{2m})=
\nonumber\\
 & =2\pi k \lim\limits_{L\to\infty}
\langle\sum\limits_n\delta(E-E_n)|\Psi_n(x_{1})|^{2q}
|\Psi_n(x_{2})|^{2q} \dots|\Psi_n(x_{2m-1})|^{2q} |\Psi_n(x_{2m})|^{2q}\rangle=
\nonumber\\
 &=\lim\limits_{L\to\infty}\lim\limits_{\epsilon\to
+0}{\displaystyle \frac{2^{2qm}(qm-1)!\,\epsilon^{2qm-1}k}{(2qm-2)!}}
\langle |G(x_{1},x_{2}|E+i\epsilon)..
G(x_{2m-1},x_{2m}|E+i\epsilon)|^{2q}\rangle=
\nonumber\\
 &=2\pi k \left\langle\left(\prod\limits_{j=1}^{m} u^{2q}(x_{2j-1})\tilde
u^{2q}(x_{2j})\right)\left(W^\prime(E)\right)^{-2qm+1}
\delta(W(E))\right\rangle=
\\
 &=2\pi k \left\langle\left(\prod\limits_{j=1}^{m} u^{2q}(x_{2j-1})
u^{2q}(x_{2j})\right)
\left(\int\limits_{-L}^{L}u^2(y)\,dy\right)^{-2qm+1}
\left(u^\prime(L)\right)^{-1}
\delta(u(L))\right\rangle\approx
\nonumber\\
 &\approx
\left\langle \prod\limits_{j=1}^{2m}
\left(v_{1}(x_{j})v_{2}(x_{j})
\right)^{q}
\left(
\int\limits_{-L}^{L}\!\!v_{1}(y)v_{2}(y)dy\right)^{-2qm+1}
\left(v_{1}(L)v_{2}(L)\right)^{-1}\!
\right\rangle=
\nonumber\\
 &\;\;\;\;\;\;
={\displaystyle \frac{N^\prime}{(2\alpha)^{2qm-1} (2qm-2)!}}
\exp\left(-{\displaystyle \frac{\alpha L}{4}}\right)\times
\nonumber\\
&\times\int\!\!\!
\int\limits_{-\infty}^{+\infty}d\sigma d\sigma^\prime
e^{(\sigma+\sigma^\prime)/2}\!\!\!
\int\limits_{\xi(-L)=\sigma^\prime,\xi(L)=\sigma}\!\!\!{\cal D}\xi
\exp \left(-\frac{1}{2\alpha}\int\limits_{-L}^{L}dt
\left(\dot{\xi}^2+e^{-\xi}\right)\right)
\exp \left(-\sum\limits_{j=1}^{2m} q\xi(x_{j})\right)=
\nonumber\\
 &={\displaystyle \frac{1}{(2\alpha)^{2qm-1} (2qm-2)!}}
\exp\left(-{\displaystyle \frac{\alpha (x_{2m}-x_1)}{8}}\right)\times
\nonumber\\
&\times\langle
\Upsilon_{0}(\xi)e^{-q\xi}|e^{-(x_2-x_1)\hat{H}}
e^{-q\xi}e^{-(x_3-x_2)\hat{H}}e^{-q\xi}\dots e^{-q\xi}
e^{-(x_{2m}-x_{2m-1})\hat{H}}
|\Upsilon_{0}(\xi)e^{-q\xi}\rangle=
\nonumber\\
&=\left(\frac{\alpha}{8}\right)^{2mq-1} 2^{4(q+m)-6}
{\displaystyle \frac{\pi^{m+1}}{\Gamma (2mq-1)\left(\Gamma (2q)\right)^2}
\left(\frac{\Gamma(q)}{\Gamma(q+1/2)}\right)^{2m-2}}
\exp\left(-{\displaystyle \frac{\alpha (x_{2m}-x_1)}{8}}\right)\times
\nonumber\\
&\times\int\limits_{0}^{\infty}\!\!.\int\limits_{0}^{\infty}
\prod\limits_{j=1}^{2m-1}d\nu_{j}\,
\exp\left(-\frac{\nu_{j}^2\alpha }{2}\Delta x_{j}\right)
\nu_{j}\,\sinh 2\pi\nu_{j}\,P^{(q)}(\nu_1)P^{(q)}(\nu_{2m-1})\!
\prod\limits_{s=1}^{2m-2}\!Q^{(q)}(\nu_{s},\nu_{s+1}),\nonumber
\end{eqnarray}
where $\Delta x_{j}=x_{j+1}-x_{j}$ and the functions
$P^{(q)}(\nu)$ and $Q^{(q)}(\nu,\nu^\prime)$
are defined as follows:
\begin{equation}\label{Pq-def}
P^{(q)}(\nu)=\frac{1}{\cosh^2\pi\nu}\left[(q-1/2)^2+\nu^2\right]
\prod\limits_{j=1}^{q-1} \left( (j-1/2)^2+\nu^2\right)^2,
\end{equation}
\begin{equation}\label{Qq-def}
Q^{(q)}(\nu,\nu^\prime)
=\frac{\nu^2-{\nu^\prime}^2}{\cosh 2\pi\nu-\cosh 2\pi\nu^\prime}
\prod\limits_{j=1}^{q-1} \left[1+ \frac{2}{j^2}(\nu^2+{\nu^\prime}^2)
+\frac{1}{j^4}(\nu^2-{\nu^\prime}^2)\right].
\end{equation}
The expression for $p^{(q,1)}_{E}(x_{1}, x_{2})$ can be obtained
from (\ref{p-qm})by the formal substitution 1 for the product
from $s=1$ to $s=2m-2$
and by putting $m=1$ in the remaining integral.

As one of possible applications of
the formulae (\ref{p-qm})-(\ref{Pq-def}) let us consider the dispersion
of the sizes of localized wave functions. It is seen from
(\ref{p-qm}) that the distant exponential asymptotics of the
probability distributions does not fluctuate. On the other hand,
it would be natural to define the wave packet size $R_{E}$ as
some integral property. For example, let us define
$R_{E}$ as:
\begin{equation}\label{size-def}
R_{E}^{-1}=\frac{4}{3}\int\limits_{-L}^{L} dx\,|\psi(x)|^4.
\end{equation}
The coefficient 4/3 cancels the mean value of the fast oscillating
factor $\sin^4(kx+\delta)$
(see (\ref{u-sinus})). Then we have:
\begin{equation}\label{size-rez}
\langle R_{E}^{-1}\rangle=
\frac{4}{3\rho(E)}\int\limits_{-L}^{L} dx\,p_E(x,x)=
\frac{2\alpha}{9}=\frac{4}{9l}.
\end{equation}
Here $\rho(E)$ is the density of states at large
$E$:
\begin{equation}\label{den-states}
\rho(E)=\frac{L}{\pi k}.
\end{equation}
The expectation value of the square of $R_{E}^{-1}$ can be found
from the correlation function
$\langle |\psi(x)\psi(x^\prime)|^4\rangle$:
\begin{equation}\label{size-sq}
\langle R_{E}^{-2}\rangle=
\frac{32}{9\rho(E)}\int\limits_{-L}^{L} dx\,
\int\limits_{-L}^{x} dx^\prime\,p_E^{(2,1)}(x^\prime,x).
\end{equation}
Using the explicit expression for $p_E^{(2,1)}(x^\prime,x)$:
\begin{equation}\label{p(2,1)}
p_E^{(2,1)}(x^\prime,x)=\frac{\pi \alpha^3}{576}
e^{-\alpha(x-x^\prime)/8}
\int\limits_{0}^{\infty} d\nu
\exp\left(-\frac{\alpha\nu^2}{2}(x-x^\prime)\right)
\frac{\sinh\pi\nu}{\cosh^3\pi\nu} \nu\left(\frac{9}{4}+\nu^2\right)^2
\left(\frac{1}{4}+\nu^2\right)^4,
\end{equation}
and evaluating the integrals over $dx^\prime$ and $d\nu$ we obtain:
\begin{equation}\label{size-sq-rez}
\langle R_{E}^{-2}\rangle\approx 0.23
\frac{1}{l^2}.
\end{equation}
With (\ref{size-rez}) it gives us the mean square relative
dispersion of $R_{E}^{-1}$:
\begin{equation}\label{size-disp}
\frac{\langle R_{E}^{-2}\rangle-\langle R_{E}^{-1}\rangle^2}
{\langle R_{E}^{-2}\rangle}
\approx 0.13.
\end{equation}

\section{Mean current in 1D mesoscopic ring with the magnetic
 flux $\Phi$}

Let us consider a one dimensional metal ring in transverse magnetic
field. The expectation value of the current operator for one electron
stationary state becomes non-zero and the energy receives
$T$
-odd term. Then the Fermi levels for left and right directions of
the mean velocity turn out to be shifted one about another.
As a result, a persistent current flows along this ring in the ground
state \cite{Bu}, \cite {BuBu}.

For a complete treatement of this physical system one should take into
account many-body effects such as electron-electron Coulomb
interaction (see e.g. \cite {G}),
fluctuations of the chemical potential \cite {Alt}
etc. I solve here only the one-particle problem and I compute
the mean current $I$ corresponding to a one-electron state
on the Fermi level.
There are arguments ( \cite{Kop},\cite{Che}) that the total current
is close to $I$ but, of course, further investigations are needed.

We will assume the ring size $2L$ to be
comparable the mean free path.
Then the localization effects do not lead to
the total suppression of $I$, but $I$ is rendered to be a nontrivial
function of the magnetic field (see below).
(The case of ordered inhomogeneous conductor has been
considered in the paper \cite{Sreb}).

There exists the gauge by which the wave function of an
electron in the ring with the magnetic flux $\Phi$ obeys the
boundary condition:
\begin{equation}\label{psi-ring}
\psi(L)=\exp\left(2\pi i\Phi\right)
\psi(-L),
\end{equation}
and the Hamiltonian has the
previous form (\ref{Hamiltonian}). The mean absolute value of current
corresponding to a state with energy $E$ can be represented
in the limit (\ref{Regime})
as follows \cite{By}, \cite{Dor}:
\[
I=\left\langle\frac{2\pi k}{L}\sum\limits_{n}
\delta(E-E_n) |j_n|\right\rangle,
\]
\begin{equation}\label{current-def}
j_n=-\frac{1}{2\pi}\frac{\partial E_n}{\partial \Phi}.
\end{equation}
($h=c=e=1$, the magnetic flux quantum is equal to 1.)
The condition (\ref{psi-ring}) is nonlocal, therefore, the formula
(\ref{current-def}) for $I$
cannot be rewritten in terms of functions like
$u(x),\tilde u(x)$ of Sections 1,2.
It has been shown in \cite{Dor}, however, that $I$ can be expressed
directly via the elements of the ${\cal T}$-matrix (\ref{Texp}).
It is worth noting that we have functional representation just for
them.

Indeed, by the construction, the matrix ${\cal T}\equiv{\cal T}(-L,L)$
satisfies the "unitarity" conditions:
\begin{equation}\label{T-prop}
\sigma^z{\cal T}^\dagger \sigma^z={\cal T}^{-1},
det\,{\cal T}=1.
\end{equation}
Therefore we can parametrize in the following way:
\begin{equation}\label{T-param}
{\cal T}
=\left(\begin{array}{cr}\cosh\Gamma e^{i\alpha_s},&\sinh\Gamma e^{i\beta_s}\\
              \sinh\Gamma e^{-i\beta_s},&\cosh\Gamma e^{-i\alpha_s}\\
\end{array}\right),
\end{equation}
where $\Gamma$, $\alpha_s$ and $\beta_s$ are slowly varying real
functions of $L$. The mapping of the initial data space at the point
$x=-L$ into the space of solutions of the equation
(\ref{Coshi}) at the point $x=L$ is realized in the basis
$(u^\prime\pm iku)$ by the transfer-matrix $
$:
\[
T=\exp(ikL\sigma^z) {\cal T} \exp(ikL\sigma^z)=
\]
\begin{equation}\label{Transfer}
=
\left(
\begin{array}{cr}\cosh\Gamma e^{i(\alpha_s+kL)},&\sinh\Gamma e^{i\beta_s}\\
           \sinh\Gamma e^{-i\beta_s},&\cosh\Gamma e^{-i(\alpha_s+kL)}\\
\end{array}\right).
\end{equation}
The condition (\ref{psi-ring})
is equivalent for the matrix $
$ to have the eigenvalue
$e^{i\theta}$, $\theta=2\pi\Phi$:
\[
det\left(T-e^{i\theta}\right)=0,
\]
or
\begin{equation}\label{T-eigen}
\tau(E)\equiv \cosh\Gamma \cos(\alpha_s+kL)=\cos\theta.
\end{equation}
The last equation defines the set of the energy $E=k^2$ values, and,
using the formula (\ref{current-def}) for $j_n$ we obtain:
\begin{equation}\label{cur-eval}
\sum\limits_{n}\delta(E-E_n) |j_n|=
\sum\limits_{n}\delta(E-E_n)
\left|\frac{\sin\theta}{\tau^\prime
(E)}\right|
=\delta\left(\tau(E)-\cos\theta\right)
\left|\sin\theta\right|.
\end{equation}
We can get rid of the $\delta$ - function in much the same manner
as in the previous section. Averaging over the interval
$\Delta L$ ($1/k\ll\Delta L \ll l$) of the (half-) circumferences $L$
we find:
\[
I=2\langle\frac{\pi k}{L}
\delta\left(\tau(E)-\cos\theta\right)\rangle
\left|\sin\theta\right|\approx
\left\langle\frac{2}{\sqrt{\sinh^2\Gamma+\sin^2\theta}}\right\rangle
\left|\frac{k}{L}\sin\theta\right|=
\]
\begin{equation}\label{I-exp}
=\left|\frac{k}{L}\sin\theta\right|
\frac{2}{\sqrt{\pi}}
\left\langle\int\limits_{-\infty}^{+\infty}d\mu\,
\exp\left(-\mu^2(\sinh^2\Gamma+\sin^2\theta)\right)\right\rangle.
\end{equation}
It is important that $\sinh^2\Gamma$ can be rewritten so that
the direct analytical continuation from the surface
$\Sigma$ becomes possible:
\begin{equation}\label{sh-reprezen}
\sinh^2\Gamma=(1,0){\cal T}^t s^{-}{\cal T}
\left(\begin{array}{c}0\\1\end{array}\right).
\end{equation}
Here the sign $t$ denotes, as usually, transposition, and complex
conjugation does not enter this formula. For
$\sinh^2\Gamma$ to have the most simple form in terms of the fields
$\rho$,$\psi^\pm$ the field
$\psi^{-}$ should obey zero-valued initial condition:
\begin{equation}\label{Inc-zero}
\psi^-(-L)=0.
\end{equation}
Then substituting (\ref{Anzatz}) into
(\ref{sh-reprezen}) we obtain:
\begin{equation}\label{sh-exp}
\sinh^2\Gamma=
\psi^{-}(L)\int\limits_{-L}^{L}dt\,\psi^+(t)
\exp\left(-i\int\limits_{t}^{L}\rho d\tau\right).
\end{equation}
The right-hand side of (\ref{sh-exp})
is bilinear in the fields $\psi^\pm$ but it is non-local.
It is convenient to start from the transformation of the
Hubbard-Stratonovich kind:
\[
I=
2\left|\frac{k\sin\theta}{\sqrt{\pi}L}\right|
\int\limits_{-\infty}^{+\infty}d\mu\,
\exp\left(-\mu^2\sin^2\theta\right)
\left\langle\exp\left\{-\mu^2\psi^{-}(L)\int\limits_{-L}^{L}dt\,\psi^+(t)
\exp\left(-i\int\limits_{t}^{L}\rho d\tau\right)\right\}\right\rangle=
\]
\begin{eqnarray}\label{linearization}
&=2\left|\frac{k\sin\theta}{\sqrt{\pi}L}\right|
\int\limits_{-\infty}^{+\infty}d\mu
\int dz\,dz^*
\exp\left(-\mu^2\sin^2\theta-|z|^2\right)\times
\nonumber\\
&\times\left\langle
\exp\left\{-i\mu z \psi^{-}(L)
-i\mu z^*\int\limits_{-L}^{L}dt\,\psi^+(t)
\exp\left(-i\int\limits_{t}^{L}\rho d\tau\right)\right\}\right\rangle.
\end{eqnarray}
turning the exponent to the linear in $\psi^\pm$ combination.
Retracing the same path (\ref{Bos-Action})-(\ref{Jacrot})
as in the previous section, and introducing the variable
$\xi(t)$ in a manner like (\ref{xi}):
\[
\dot{\xi}=-2(1+4a)\eta+4ia\rho ,
\]
\begin{equation}\label{-xi}
\xi(L)=0
\end{equation}
\[
{\cal D}\rho{\cal D}\eta=\hbox{const}\,{\cal D}\rho{\cal D}\xi
\]
we bring the ${\cal D}\rho{\cal D}\chi $-integration to the
Gaussian form. Performing it we find:
\begin{eqnarray}\label{I-int}
&I=\exp(-\frac{\alpha L}{4})
\left|\frac{k}{L}\sin\theta\right|
\frac{2}{\pi^{3/2}} N^\prime
\int\limits_{-\infty}^{+\infty}d\mu\,
\int dz\,dz^*
\exp\left(-\mu^2\sin^2\theta-|z|^2\right)\times
\nonumber\\
&\times
\int\limits_{\xi(L)=0}{\cal D}\xi
\exp \left(-\frac{1}{2\alpha}\int\limits_{-L}^{L}\,dx
\left(\dot{\xi}^2+\alpha^2\mu^2|z|^2 e^{-\xi}\right)\;
-\frac{\xi(-L)}{2}\right).
\end{eqnarray}
Let us change $\mu$ for the integration variable $\sigma$:
\begin{equation}\label{mu-si}
\alpha^2\mu^2|z|^2=e^{-\sigma},
\end{equation}
and shift the trajectory $\xi(x)$ by $-\sigma$:
\[
\xi(x)\rightarrow\xi(x)-\sigma.
\]
Then the integral over $(z\,z^*)$-plane is calculated exactly and
we obtain the representation of
$I$ as the matrix element:
\begin{equation}\label{I-mtrx}
I=\exp(-\frac{\alpha L}{4})
\left|\frac{k}{\sqrt{\pi}\alpha L}\sin\theta\right|
\langle\Upsilon_{2}(\xi)|\exp(-2L\hat{H})|\Upsilon_{1}(\xi)\rangle,
\end{equation}
Here $\hat{H}$ is defined in (\ref{Ham-eff}), and
the functions $\Upsilon_{1,2}(\xi)$ have the following forms:
\begin{equation}\label{Ups-1}
\Upsilon_{1}(\xi)=\exp\left(-\frac{\xi}{2}\right)
\end{equation}
and
\begin{equation}\label{Ups-2}
\Upsilon_{2}(\xi)=
\exp\left\{-\frac{2}{\alpha}\exp\left(-\frac{\xi}{2}\right)
|\sin\theta|\right\}.
\end{equation}
Using the expansion in terms of the complete set (\ref{Eigenfunctions})
and the integral representation for
$K_{2i\nu}(y)$ we find the explicit formula for the current
$I$: ($l^{-1}=\alpha/2$)
\begin{equation}\label{current-rez}
I=\sqrt{\frac{8}{\pi}}
\frac{k}{L}\left(\frac{l}{L}\right)^{1/2}
\exp\left(-\frac{L}{2l}\right)
\sin^2\theta\int\limits_{-\infty}^{+\infty}
\frac{dt\,\cosh t}{\sin^2\theta+\sinh^2t}\exp\left(-\frac{l}{2L}t^2\right)
\end{equation}
When $\sin^2\theta$ runs from 0 to 1, the value of $I$
increase monotonically from 0 to $I_{max}$.
The first order of the Taylor series in $\theta$
agrees with the result of
\cite{Dor}. It is worth noting that the parameter
$l/L$ determines not only the absolute value of
$I$, but its dependence on the magnetic field as well:
\begin{equation}\label{current-dep}
\frac{I}{I_{max}}=
\frac{\sin^2\theta}{\int\limits_{-\infty}^{+\infty}
\frac{dy}{\cosh y}\exp\left(-\frac{l}{2L}y^2\right)}
\int\limits_{-\infty}^{+\infty}
\frac{dt\,\cosh t}{\sin^2\theta+\sinh^2t}\exp\left(-\frac{l}{2L}t^2\right)
\end{equation}
This formula allows us, in principle, to find $l/L$ from
the run of the experimental curve $I(\theta)$.
Finishing this section I would like to emphasize that all
its formulae are exact in the limit (\ref{Regime}) and no
analogue of the "non-resonance" terms have appeared.

\section{Conclusion}

Let us suppose for the random potential
$U(x)$ to have the finite correlation
length $\kappa^{-1}$:
\begin{equation}\label{Noise-corr}
<U(x)U(x^\prime)>=\frac{1}{2}D\kappa\exp\left(-
\kappa|x-x^\prime|\right).
\end{equation}
We can take it into account in the limit
$1\ll\kappa l$ by the renormalization of the parameter $\alpha$:
\begin{equation}\label{al-renor}
\alpha_{\kappa}=\frac{\alpha_{0}}{1+4k^2/\kappa^2}.
\end{equation}
It should be noted that this renormalization may be sufficient
since the inequalities $1\ll\kappa l$ and $k/\kappa\sim 1$
can take place simultaneously.

Indeed, the correlator (\ref{Noise-corr}) corresponds
to the following measure of the
integration over the fields $\zeta^{\pm}$:
\begin{equation}\label{measure-corr}
{\cal D}\zeta^{\pm}
\exp\left\{-\frac{2}{\alpha}\int\limits_{-L}^{L}\left(\frac{1}{\kappa^2}
|\dot{\zeta}|^2+2i\frac{k}{\kappa^2}(\dot{\zeta^+} \zeta^-
-\zeta^+\dot{\zeta^-})+
\left(1+\frac{4k^2}{\kappa^2}\right)|\zeta|^2\right)\,dx\right\}.
\end{equation}
On performing the "bosonisation" and passing to the
variable $\xi$ we obtain some effective action $S_{eff}(\xi)$.
The terms with derivatives
$\dot{\zeta^\pm}$ in the exponent of (\ref{measure-corr})
would produce the terms of $S_{eff}(\xi)$ containing
derivatives as well. The localization length
$l$ is the only parameter of the length dimension occurring in the
unperturbed problem.
Therefore the contributions of those "non-markovian" terms would
be suppressed by powers of the quantity
$(\kappa l)^{-1}$. Neglecting them we come to the formula
(\ref{al-renor}).

The variable $\sigma^\prime$ appeared in (\ref{Density-Integral}),
(\ref{Shift}) could be considered as the global order parameter
corresponding to the localization. In fact, the non-zero value of the
correlator
$p_E(x,x^\prime)$ in the thermodynamic limit is the consequence of the
following relationship:
\begin{equation}\label{order-par}
\lim\limits_{L\to\infty}\frac{\sigma^\prime}{\alpha L}=
\frac{1}{2} > 0.
\end{equation}
On the other hand, the quantity $\lambda = \exp(-\sigma^\prime) $
is conjugated to the wave function norm. Then the inequality
(\ref{order-par}) corresponds to the exponential in the average
increase of the functions
$u$, $\tilde u$ (\ref{Coshi}).

It is worth noting that the large-scale behaviour
of wave functions is governed by some averaged characteristics of
the potential $U(x)$. In the perturbation theory framework they
appear as some infrared singular integrals. The integrands are
multipoint products of potential with fast-oscillating exponentials.
Thus, if the quantity
\begin{equation}\label{l-cond}
b_{\Delta}(x,d)=
\int\limits_{x-\Delta}^{x+\Delta}dx^\prime\,
\exp\left(ik(x-x^\prime)\right) U(x) U(x^\prime)
\end{equation}
starting from some $\Delta$
\[
k^{-1}\ll\Delta\ll l
\]
becomes independent on $x$, the Abrikosov-Ryzhkin model with
some effective $a$ and $\alpha$ can be used to investigate
the properties of the wave functions.

A method similar to that presented here allows one to derive
path integral representation for any averaged combination of Green
functions at arbitrary energy
$E$ \cite{KolMi}. Unfortunately, I succeeded in computing such
path integrals in some of the simplest cases only.

\section{Acknowledgments}

I am grateful to M.Chertkov, I.B.Khriplovich,
D.L.Shepelyansky, V.V.Sokolov, O.Sushkov
and P.Silvesterov for useful conversations and notices. I am
grateful to A.Gamba and M.Martellini from Milan University
for the warm hospitality. I have benefited from one note of
M.Martellini. I would like to express my thanks to D.A.Shapiro
for his help in preparing the manuscript .
It is a please for me to acknowledge the support from the
Soros foundation.
I am thankful to the referee for the remarks allowing to
improve the text of the article.

\end{document}